\documentclass[a4paper, 11pt]{article}
\usepackage{pos}
\usepackage{graphicx} 
\usepackage{float}
\usepackage{caption}
\usepackage{subcaption}
\pdfoutput=1

\begin{document}
\title{\raggedright\textbf{Quantum Phase Transition of Non-Hermitian Systems
using Variational Quantum Techniques}}
\author*[1]{James Hancock} 
\author[1]{Matthew Craven} 
\author[1]{Craig McNeile} 
\author[1]{Davide Vadacchino} 

\affiliation[1]{University of Plymouth, Centre for Mathematical Sciences, Plymouth} 

\emailAdd{james.hancock@plymouth.ac.uk} 
\date{}

\FullConference{The 41st International Symposium on Lattice Field Theory (LATTICE2024)\\
 28 July - 3 August 2024\\
Liverpool, UK\\}

\abstract{
The motivation for studying non-hermitian systems
and the role of $\mathcal{PT}$-symmetry is discussed.
We investigate the use of a quantum algorithm to find the eigenvalues and eigenvectors of non-Hermitian Hamiltonians, with applications to quantum phase transitions. We use a recently proposed variational algorithm. The systems studied are the transverse Ising model with both a purely real and a purely complex transverse field.
}

\maketitle

\section{Introduction}

Quantum computers offer the possibility of simulating non-perturbative quantum field theories 
in some regions of parameter space
that were previously inaccessible using standard lattice field theory techniques, based on Monte Carlo Methods. See~\cite{Funcke:2023jbq,DiMeglio:2023nsa} for recent reviews of using quantum computers
to solve problems in particle physics.
Ultimately, we are interested in using quantum computers for systems affected by a sign problem, such
as a field theory with a background electric field~\cite{Yamamoto_2021}.
As a first step, we study the features of a possible quantum phase transition (QPT) due to a change in an imaginary magnetic field. 
QPTs are distinct changes in the form of the ground state of a system at zero temperature, driven by quantum fluctuations, due to a change in an 
external parameter such as the magnetic field~\cite{QPTBOOK}.
The introduction of a complex magnetic field makes the Hamiltonian non-Hermitian. This preliminary study will allow us to develop some of the tools required for simulations of systems with a sign problem.

Non-Hermitian quantum mechanics is a framework for computing the dynamics of
open quantum systems, such as quantum 
channels~\cite{vijaywargia2024quantumclassicalcorrespondencequantumchannels}. This connection suggests that quantum resources themselves could be better understood and optimized through the study of non-Hermitian systems. Non-Hermitian quantum mechanics has applications in the field of optics \cite{NHOPTICS}, topological materials in condensed matter \cite{Okuma:2022bnb}
and other areas \cite{ghangas2024coherentcontrolnonreciprocaloptical,zhang2024skineffectnonhermitiansystems}.
See \cite{Moiseyev2011} for an extensive review about non-Hermitian quantum mechanics.

A special case of non-Hermitian quantum mechanics are
Hamiltonians with $\mathcal{PT}$-symmetry. Although the Hamiltonian is non-Hermitian, it may have real eigenvalues.
$\mathcal{PT}$-symmetric quantum mechanics is particularly useful in optical systems with balanced loss and gain, such as lasers \cite{PTLASERS}. Such systems exhibit two distinct phases, defined by the breaking (or not) of $\mathcal{PT}$-symmetry. The transition between these phases occurs at specific parameter values known as exceptional points. Physics near exceptional points has promise in improving experiments, such as efficiency for energy harvesting \cite{Fernández-Alcázar2021} and sensitivity of the detection of splittings of resonant frequencies \cite{PhysRevLett.112.203901}. There are still some potential conceptual issues with the definition of the transfer matrix for $\mathcal{PT}$-symmetric quantum mechanics~\cite{Alexandre:2023afi}. Bernard	and Savage~\cite{Bernard:2001wh} have developed lattice
methods for the numerical simulations of $\mathcal{PT}$-symmetric Hamiltonians. For a more detailed discussion on $\mathcal{PT}$-symmetric quantum mechanics 
see~\cite{Bender_2005,Bender_2007}.

Classical Monte Carlo-based methods are unsuitable for simulating non-Hermitian systems
because of the sign problem caused by the complex eigenvalues~\cite{Troyer_2005}. 
Therefore in this work, we investigate algorithms for quantum computers to compute the eigenvalues of
 non-Hermitian systems. We note that our initial test cases do not respect $\mathcal{PT}$-symmetry.
 
 There is a connection between the quantum Ising model in $d$ dimensions and an anisotropic classical
Ising model in $d+1$ dimensions. There have been a number of 
studies~\cite{matveev2008properties,Azcoiti:2016hpa} of the classical Ising
model with complex magnetic fields in two dimensions. The phase structure of the anti-ferromagnetic classical Ising model in 2D with complex magnetic field is rich. This motivates us to study the quantum phase structure of the anti-ferromagnetic 1D quantum Ising model with a complex magnetic field using quantum computing techniques in future work.

\section{Introduction to Non-Hermitian Quantum Mechanics}
In quantum mechanics, systems are typically described by a Hamiltonian $\hat{H}$, which is usually Hermitian (self-adjoint). The Hermiticity of $\hat{H}$ ensures several important physical properties: real energy eigenvalues, conservation of probability (norm), and a complete orthonormal set. In this study, we explore the effects of introducing non-Hermitian terms to the Hamiltonian. Such a Hamiltonian can be expressed as
\begin{equation}\label{eq:nHHamil}
    \hat{\mathcal{H}} = \hat{H} + i\hat{\eta},
\end{equation}
where both $\hat{H}$ and $\hat{\eta}$ are Hermitian operators. As they are Hermitian, we can write these sub-Hamiltonians as a sum over Pauli string matrices - suitable for use with quantum algorithms. We note that any general matrix may be expressed as this, known as the Toeplitz decomposition \cite{Horn2013Matrix}.

\section{Systems Studied}  \label{sec:IsingModel}

In this work, we investigate the effectiveness of this method over two systems.
The first is the standard Ising chain, defined by the Hamiltonian
\begin{equation}\label{eq:TIM}
    \hat{H} = -\sum_{j=0}^{n-1}\left[\sigma_z^j\sigma_z^{j+1} + \Gamma\sigma_x^j\right]
\end{equation}
with periodic boundary conditions. This system has a 
quantum phase transition at $\Gamma=1$  in the infinite volume limit~\cite{Sachdev2011}.

The second model we studied is the Ising model with an imaginary transverse field, for which we
consider the Hamiltonian
\begin{equation}\label{eq:IM}
    \hat{\mathcal{H}}_{I} = -\sum_{j=0}^{n-1}\left[\sigma_z^j\sigma_z^{j+1} + i\Gamma_{I}\sigma_x^j\right],
\end{equation}
where $\Gamma_I$ controls the non-Hermicity of the system. Clearly, we can write this in the form given in Eq. (\ref{eq:nHHamil}). This Hamiltonian is non-Hermitian, and does not exhibit $\mathcal{PT}$-symmetry.
There has been a previous study, using numerical diagonalization, 
of a non-Hermitian transverse field Ising model (see Ref. \cite{Starkov_2023}), with a staggered longitudinal magnetic field that is $\mathcal{PT}$-symmetric. 

For quantum phase transitions the temperature is zero and phase transitions are caused by
quantum fluctuations. The groundstate $|\phi_0\rangle$ of the Hamiltonian is computed
and used to estimate quantities, such as the magnetization in the transverse direction, defined by
\begin{equation}\label{eq:Mx}
   \langle\hat{M}_x\rangle := \langle \phi_0|\hat{M}_x|\phi_0\rangle = \frac{1}{n}\sum_{j=0}^{n-1}\langle\phi_0|\sigma_x^j|\phi_0\rangle \; .
\end{equation}

For the transverse Ising model, we also calculated the susceptibility in the $\sigma_x$-direction, 
defined by
\begin{equation}\label{eq:chi}
    \chi_x \approx \langle \hat{M}_x^2\rangle - \langle \hat{M}_x\rangle^2,
\end{equation}
where we can only consider this an approximation due to working at zero temperature. We use this as an order parameter to extract physics in the thermodynamic limit.

\section{Variational Quantum Algorithm for Finding Non-Hermitian Eigenvalues}\label{sec:VQA}
An important algorithm for computing eigenvalues of Hermitian operators in the NISQ (noisy intermediate-scale quantum) era of quantum computing is the variational quantum eigensolver (VQE)~\cite{Tilly:2021jem}. 
The standard VQE will not work with non-hermitian
operators.
Xie, Xue, and Zhang have developed~\cite{Xie_2024}
a variational algorithm to compute the eigenvalues of non-Hermitian operators. 
Their methodology provides a way to massage Hamiltonians in the form given in Eq.~(\ref{eq:nHHamil}) into a Hermitian form that enables its use on a quantum computer to evaluate expectation values. We introduce a parameter $E\in\mathbb{C}$, and define two new operators
\begin{equation}
    \hat{M}^+(E;\hat{\mathcal{H}}) = \left(\hat{\mathcal{H}}^\dagger - E^*\right)\left(\hat{\mathcal{H}}-E\right) \quad \text{and} \quad \hat{M}^-(E;\hat{\mathcal{H}}) = \left(\hat{\mathcal{H}}-E\right)\left(\hat{\mathcal{H}}^\dagger - E^*\right).
\end{equation}
The operators $\hat{M}^+(E;\hat{\mathcal{H}})$ and
$\hat{M}^-(E;\hat{\mathcal{H}})$ are both Hermitian and positive semi-definitive.

We can use a quantum computer to find the expectation value of the operator $\hat{M}^+$ with the parameterized state in the VQE. In general, even for simple non-Hermitian Hamiltonians, this expression contains all three $\sigma_x$, $\sigma_y$ and $\sigma_z$ Pauli operators acting on arbitrary qubits, and thus a fully or over-expressive ansatz is used. We will denote the parameterized state by $|\phi(\theta)\rangle = G(\theta)|0\rangle^{\otimes n}$, where $\theta$ is a vector of parameters and $G$ is a circuit created from single-qubit rotational gates and two-qubit $CNOT$s.

Calculating the expectation value of this operator in the parameterized state leads to two cost functions
\begin{equation}
    C^+(\theta, E;\hat{\mathcal{H}}) = \left\langle \phi(\theta)\left|\hat{M}^+\right|\phi(\theta)\right\rangle \quad \text{and} \quad C^-(\theta,E;\hat{\mathcal{H}}) = \left\langle \phi(\theta)\left|\hat{M}^-\right|\phi(\theta)\right\rangle.
\end{equation}
It can be seen that $C^+ =0$ when $|\phi(\theta)\rangle$ and $E$ for a right-hand eigenpair of $\hat{\mathcal{H}}$ and $C^- =0$ when they form a left-hand eigenpair. We only investigated minimizing $C^+$.

In the original work~\cite{Xie_2024} a two-stage gradient descent-based optimization strategy was used, where the steps depend on the particular eigenpair being searched for. We have explored adaptations of this, including using more sophisticated optimization techniques such as Adam~\cite{kingma2014adam}. 
We have found that a slightly modified version of the original strategy is most effective, wherein another stage is used for the parameters. We focus on achieving an acceptable level of accuracy in a noisy environment with less concern for the speed of solving.


A quantum computer has been used to simulate the thermal evolution of a transverse field Ising model~\cite{Cervera_Lierta_2018}. Another formalism to modify the VQE for non-Hermitian matrices has recently been developed~\cite{Zhao:2023fyz}. There have been other proposals to modify the variational method to deal with non-Hermitian matrices~\cite{detar1979variational,Kraft:2016rff}.

\section{Results}

We used two methods to study the quantum phase transitions in the two models.
Originally we investigated using the \texttt{qiskit} software from IBM to run the quantum simulations. Unfortunately, it was too slow.
Hence, a state vector representation of the quantum computer was used to estimate the
groundstate of the models using the non-Hermitian VQE.
Due to resource limitations, we could only use five spin sites with five qubits.
At each inner product (measurement) calculation, a random variable  uniformly distributed in the range $(-\epsilon, \epsilon)$, $\epsilon = 0.04$. By running experiments on the same system using \texttt{qiskit}, this is comparable to using $\sim 1000$ shots for five qubits but does not capture the complexity of true noise, but means we are still working in a noisy enviroment. 

We also used the  C++ `Eigen' library\cite{eigenweb} in C++ to directly 
compute the eigenpairs as a comparison. This is the classical method we compare against the
quantum computing method. The C++ implementation allowed us to run for more iterations on larger systems.

\begin{figure}[]
    \centering
        \begin{subfigure}[t]{0.49\linewidth}
        \centering
        \includegraphics[width=0.975\linewidth]{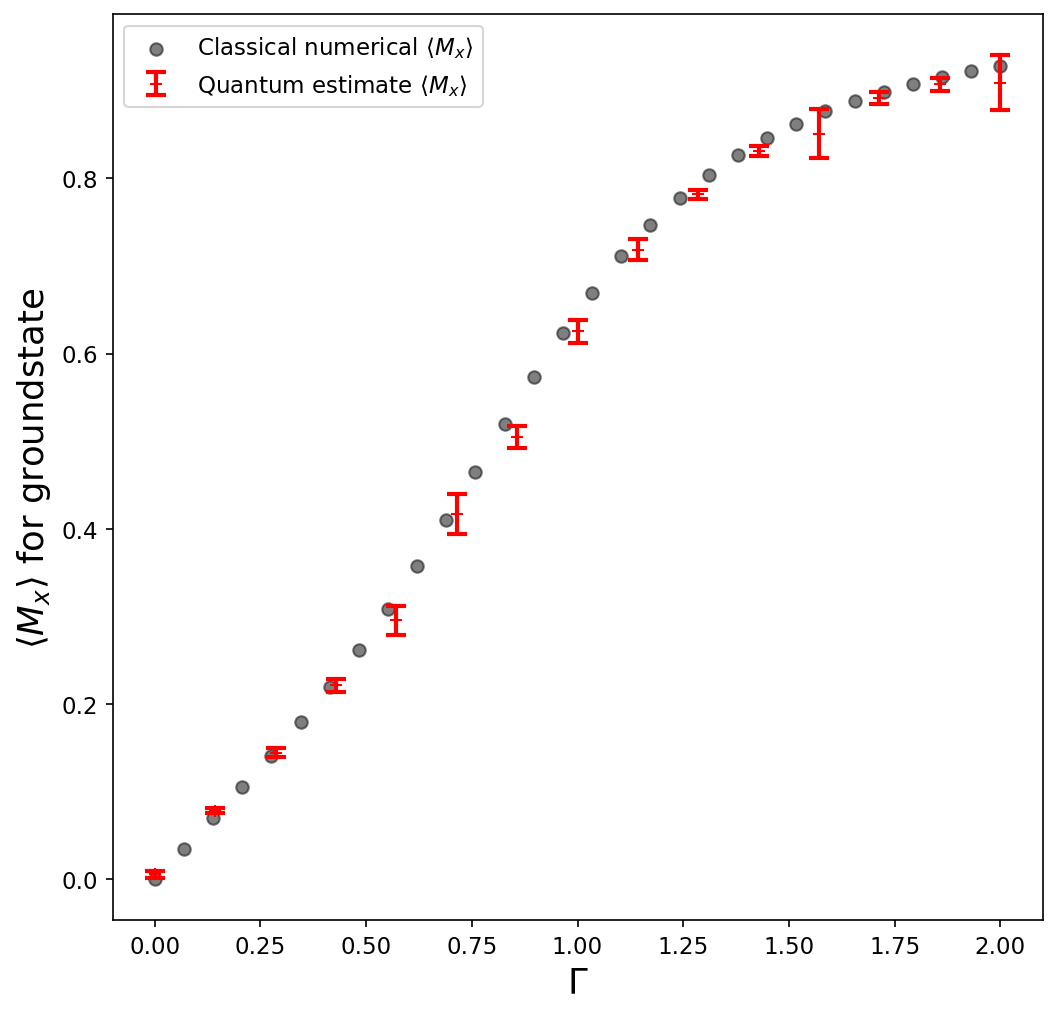}
        \caption{$\langle\hat{M}_x\rangle$ (Eq. \ref{eq:Mx}) over a change in real magnetic field ($\Gamma$) in the transverse Ising model (Eq. \ref{eq:TIM}).}
        \label{fig:CTIM3}
    \end{subfigure}
    \hfill
    \begin{subfigure}[t]{0.49\linewidth}
        \centering
        \includegraphics[width=\linewidth]{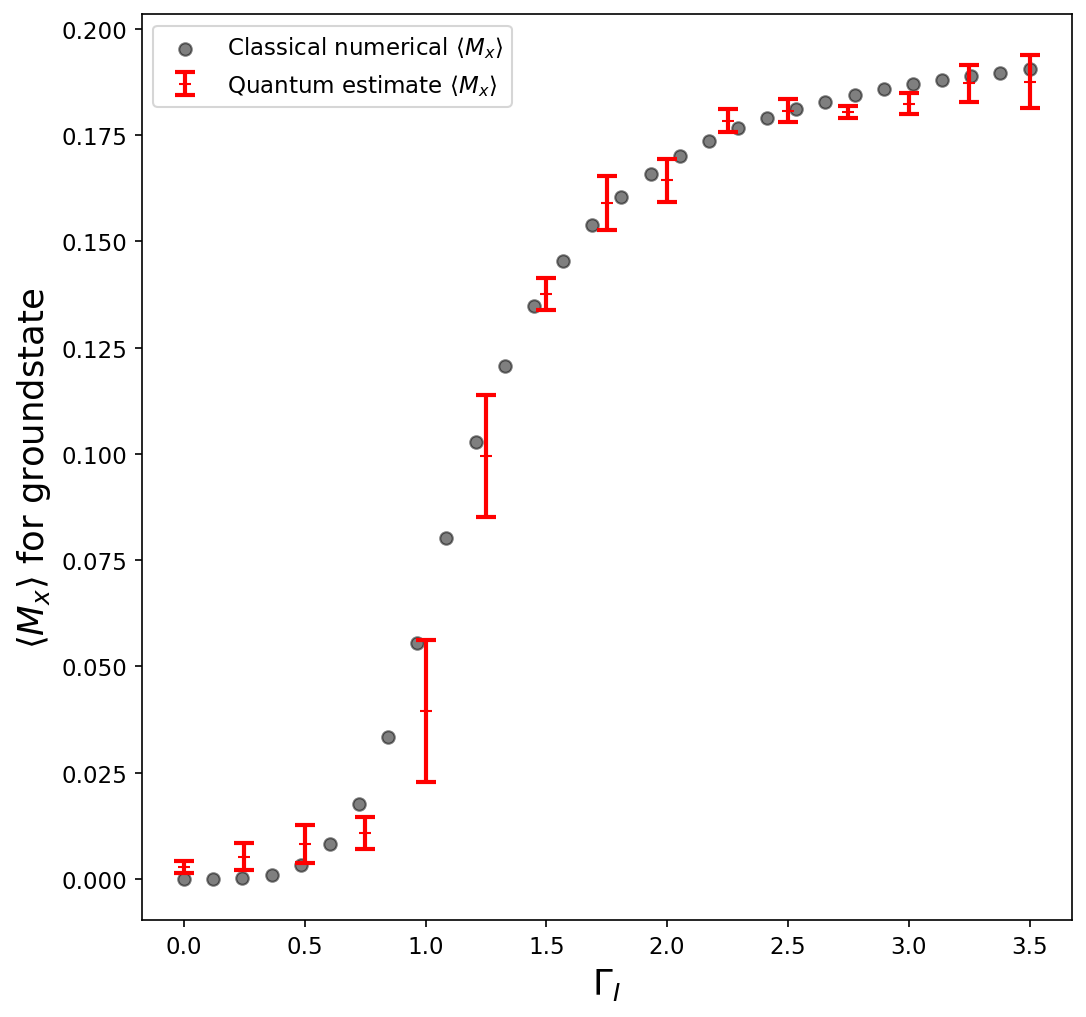}
        \caption{$\langle\hat{M}_x\rangle$ (Eq. \ref{eq:Mx}) over a change in imaginary magnetic field ($\Gamma_I$) in the non-Hermitian transverse Ising model (Eq. \ref{eq:IM}).}
        \label{fig:CTIM1}
    \end{subfigure}%
    \caption{$\langle\hat{M}_x\rangle$ over a change in magnetic field for the transverse Ising model and non-Hermitian transverse Ising model with 5 spins.}
    \label{fig:CTIM_Comparison}
\end{figure}

The $\sigma_x$--magnetization (as defined in Sec. \ref{sec:IsingModel}) is used for the ground state as the order parameter for measuring the QPT in each system. 
 The results for each system for $\langle\hat{M}_x\rangle$ as a function of magnetic field  
are in Fig.~\ref{fig:CTIM_Comparison} using the method from Sec. \ref{sec:VQA}.

$\langle\hat{M}_x\rangle$ for five spin transverse Ising models with real magnetic field and imaginary magnetic field are in Figs~\ref{fig:CTIM3} and \ref{fig:CTIM1} respectively. Good agreement is obtained between the quantum and classical (using explicit diagonalisation) algorithms. The shape of the two plots are qualitatively different, but any conclusions requires a study of the finite volume corrections.

\section{Infinite Volume Limit}

 Our implementation of the non-Hermitian quantum eigensolver was limited to five spins only due to
 computational limitations. We used exact diagonalisation to study the transverse Ising model with a large number of spins, so we could investigate the extrapolation to the infinite volume
 limit.

 We use the susceptibility in the $\sigma_x$-direction: $\chi_x$, as the parameter to extrapolate behavior at the thermodynamic limit. We choose this because it shows clear trends in the peak (growing and changing position), features which are more difficult to see in the $\langle \hat{M}_x\rangle$ curves. 
 
 \subsection{Results for the transverse Ising Model in real magnetic field}

\begin{figure}
\centering
  \centering
  \includegraphics[width=0.5\linewidth]{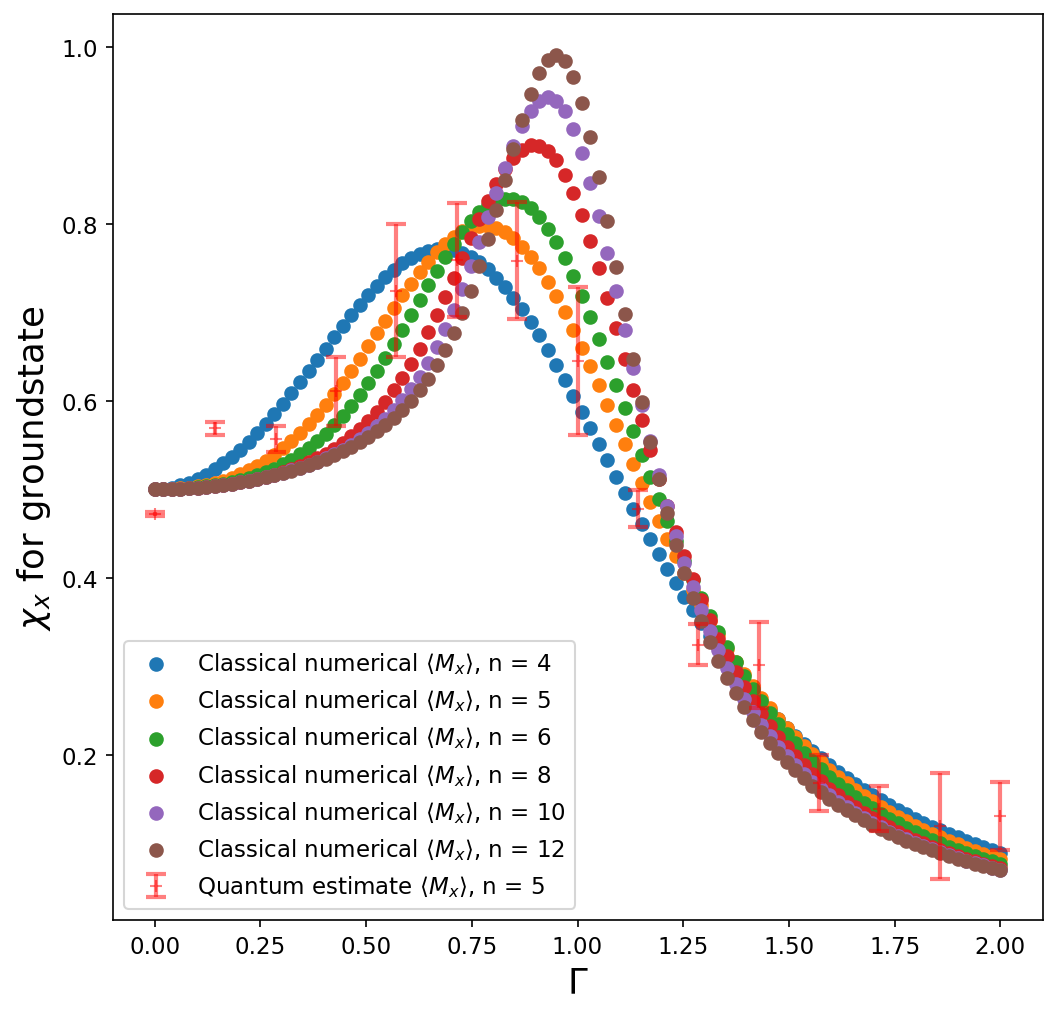}
  \caption{Magnetic susceptibility in the $\sigma_x$-direction, $\chi_x$ (Eq. \ref{eq:chi}), for the transverse Ising model (Eq. \ref{eq:TIM}), for a varying number of qubits}
  \label{fig:sub1}
\end{figure}

We can use $\chi_x$ to get a handle on the finite volume effects by studying how the peak value moves for different system sizes.
Our implementation of the quantum algorithm on the hardware we used was restricted to the maximum of 5 qubits. However the exact diagonalisation method could simulate larger systems,so the finite volume corrections could be studied.
In Fig.~\ref{fig:sub1},  $\chi_x$ is plotted for the transverse Ising model with a real magnetic field with spins between 5 and 12 obtained using exact diagonalisation. For five spins, the results from the quantum simulator are also reported.
There is good evidence for a quantum phase transition at $\Gamma$ = 1, from the peak in $\chi_x$. This is consistent with 
other studies~\cite{Sachdev2011}.

\subsection{Results for the Non-Hermitian Ising model}

\begin{figure}
\centering
  \centering
  \includegraphics[width=0.5\linewidth]{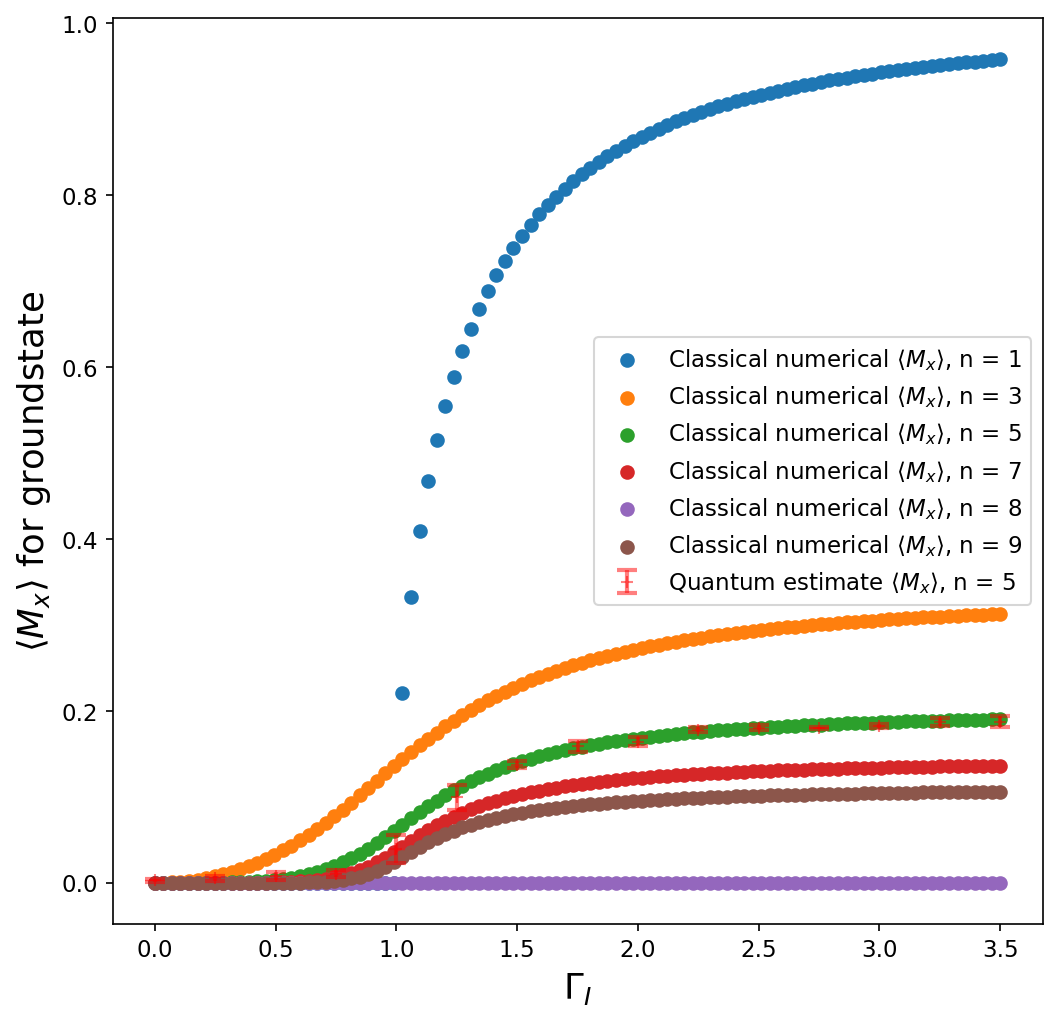}
  \caption{Magnetic charge per spin in the $\sigma_x$-direction, $\langle\hat{M}_x\rangle$ (Eq. \ref{eq:Mx}), for the non-Hermitian transverse Ising model (Eq. \ref{eq:IM}), for a varying number of qubits. }
  \label{fig:suba}
\end{figure}

We computed $\langle\hat{M}_x\rangle$ for the non-Hermitian transverse Ising model defined in Eq. \ref{eq:IM}.
The results for these were gathered using matrix/statevector representations and C++ `Eigen' methods \cite{eigenweb}. 
We did compute $\chi_x$, but it was less useful for this system.
The results for $\langle\hat{M}_x\rangle$ as a function of
$\Gamma_I$ are plotted in Fig.~\ref{fig:suba}. The functional
dependence of $\langle\hat{M}_x\rangle$ on $\Gamma_I$ is very different for an even and odd number of spins. The results for
an even number of spins show $\langle\hat{M}_x\rangle$ = 0 
for all $\Gamma_I$, but this is different to the functional form for an odd number of spins. As the number of odd spins increases 
 $\langle\hat{M}_x\rangle$ does look as though it is tending to zero as the number of spins increases, so the results will
 be consistent for even and odd spins in the thermodynamic limit.
Another study found differences in the order parameter of an antiferromagnetic transverse Ising model for small numbers of spins with different boundary conditions~\cite{He_2017}. 

As there is some evidence that for larger systems $\langle\hat{M}_x\rangle= 0$ for all $\Gamma_I$ for the groundstate, this suggests that there is no quantum phase transition. To be definitive, much larger systems should be simulated.

\section{Conclusions}

We have investigated methods to study QPTs in non-Hermitian systems by using a modified
algorithm for quantum computers to compute the eigenpairs proposed in~Ref. \cite{Xie_2024}.
We see promising indications of the ability to quantify and study the behavior of possible quantum phase transition. 
 
We are investigating GPU acceleration of our quantum simulators using CUDA-Q and other packages~\cite{stein2024cuaoanovelcudaacceleratedsimulation}.
Also, we will study the effect of noise in the quantum simulator on the final results, in addition to running on a real quantum computer.

\newpage 

\end{document}